\documentclass[reprint,amssymb, amsmath, aip,cha,twocolumn]{revtex4-1}
\usepackage{graphicx}
\usepackage{color}
\usepackage{epsfig}
\usepackage{bm}%
\usepackage[colorlinks=true,linkcolor=blue]{hyperref}%
\expandafter\ifx\csname package@font\endcsname\relax\else
 \expandafter\expandafter
 \expandafter\usepackage
 \expandafter\expandafter
 \expandafter{\csname package@font\endcsname}%
\fi
\begin{document}

\title{Theoretical study of head-on collision of dust acoustic solitary waves in a strongly coupled complex plasma}%
\author{S.Jaiswal}%
\email{surabhi@ipr.res.in}
\author{P.Bandyopadhyay}
\author{A.Sen}
\affiliation{ Institute For Plasma Research, Bhat, Gandhinagar,Gujarat, India, 382428}%
\date{\today}
\begin{abstract}
We investigate the propagation characteristics of two counter propagating dust acoustic solitary waves (DASWs) undergoing a head-on collision, in the presence of strong coupling between micron sized charged dust particles in a complex plasma. A coupled set of nonlinear dynamical equations describing the evolution of the
two DASWs using the extended Poincar\'{e}--Lighthill--Kuo perturbation technique is derived. The nature and extent of post collision phase-shifts of these solitary waves are studied over a wide range of dusty plasma 
parameters in a strongly and a weakly coupled medium. We find a significant change in the nature and amount of phase delay in the strongly coupled regime as compared to a weakly coupled regime. The phase shift is seen to change its sign beyond a threshold value of compressibility of the medium for a given set of dusty plasma parameters.
\end{abstract}
\maketitle
\section{Introduction} 
It is well known that the dynamics of dusty plasmas is very different from that of the usual two components electron-ion plasma. In 
a complex (dusty) plasma, particles ranging from nanometer to micrometer size are immersed in an ionized gaseous medium with
free-floating electrons and ions. The dust particles become charged and interact collectively, and depending upon the system parameters 
can be in the gaseous or liquid state or even arrange themselves to form an orderly crystalline structure in a solid state\cite{Ikezi, thomas_1994}. 
The complex plasma medium also
supports a variety of collective modes and nonlinear coherent structures, such as dust-ion-acoustic (DIA) waves
\cite{ma}, dust-acoustic (DA) waves \cite{rao}, dust lattice (DL) waves \cite{melands}, dust Coulomb waves \cite{N}, dust 
voids \cite{goree} and vortices \cite{law}. Among them, the DIA solitary waves, the DA solitary waves, the DL 
solitary waves, and the envelope of DIA/DA solitary waves are very important nonlinear waves which have been extensively studied both 
theoretically \cite{shuk, kour, morf, versheest, mamun}, and experimentally \cite{Pintu, samsonov, heidemann} over the past few years.
\par Interaction of solitons is an interesting phenomenon in normal fluids and plasma which has been reported by a number of
authors \cite{zabusky, tsuji, gard, xue, han, shamy, labany, moslem, versh, ghosh}. The observation that a discrete 
nonlinear system exhibits recurrent states instead of an ergodic behavior was explained by Zabusky and Kruskal \cite{zabusky}, who
realized that the nonlinear chain is described by the KdV equation in the continuum limit. Thus any deviation from KdV approximation in real systems will affect the thermalization time and break the recurrence of the
 initial state. 
 In a one dimensional system, the solitons may interact amongst themselves in two different ways. There can be an overtaking collision (for co-propagating solitons), which can be studied by the inverse scattering transform method \cite{gard} or a head-on collision (counter-propagating solitons), where the angle between the two propagation directions of the two solitons is equal to $\pi$. For a head--on collision between two solitary waves travelling from positive and negative directions,
 the important post-collision consequences to study are the phase shifts and resultant changes in their respective trajectories. Many authors have investigated the head-on collision of two solitary waves in different plasma models using the extended 
 Poincar\'{e}--Lighthill--Kuo (PLK) method \cite{xue, han, shamy, labany, moslem, versh, ghosh}. In a dusty plasma,  Xue \cite{xue} investigated the head-on collisions of dust--acoustic solitary waves in an unmagnetized dusty plasma including the dust charge variation, and the analytical phase shifts following the head--on collision were derived. They showed the variation of phase shift with different plasma parameters. Ghosh \textit{et al.} \cite{ghosh} studied the head-on collision of dust acoustic solitary waves in a four component unmagnetized dusty plasma with Boltzmann distributed electrons, non-thermal ions, and negatively charged dust grains
as well as positively charged dust grains. 
 Very recently, a couple of experimental observations \cite{Harvey, bailung2014} have been reported separately on the head-on collision of two counter-propagating solitary waves. They found that both the solitary waves pass through each other and suffer a small time delay in their propagation after the collision. In one of the observations, Harvey \textit{et al.} \cite{Harvey} found the sum of the amplitude of individual solitons is  less than that of resultant solitary amplitude whereas the Sharma et al. \cite{bailung2014} reported that the resultant amplitude is exactly equal to the sum of individual amplitudes. But in both the cases, it is observed that the solitons with higher amplitude experience longer delays. \par 
To the best of our knowledge, there has been no detailed investigation on the interaction of solitary waves in a strongly coupled dusty
 plasma.  In this paper, we investigate the head-on collision of
 dust acoustic solitary waves and deduce the phase shifts by an extended version of the PLK method considering the effects of strong
 coupling between the dust particles in an appropriate fluid model. The fluid model we adopt is based on the Generalized Hydrodynamic Equations
 which have been used in the past to study the linear \cite{kaw, mishra} and non-linear \cite{veeresha} propagation of dust acoustic waves in a strongly coupled dusty plasma. These
 studies have shown that strong coupling effects introduce additional dispersive effects in the linear propagation characteristics of DAWs through
 modifications in the compressibility and visco-elastic properties of the system. An interesting consequence is the ``turn-over" effect where beyond 
 a certain value of the strong coupling parameter the group velocity of the DAW can change sign and travel backward. A nonlinear manifestation of
 this effect should be of interest to investigate and is a major motivation for our present study. To explore this effect we have looked at the variation of the phase shift 
 that arises due to the head-on collision between two dust acoustic solitary waves, over
 a wide range of plasma and dusty plasma parameters. It is found that there is a significant change of phase shift in the strongly coupled
 regime compared to the weakly coupled regime and beyond a critical value of the compressibility there is a change in the sign of the phase shift.  \par
 The paper is organized as follows.  In the next section, we present the model equations to deduce the expression for phase 
 shifts of two solitons after making a head-on collision by taking into account strong coupling between the particles. In Sec. 
 III, we discuss our analytical results of phase shifts  in a wide range of plasma and dusty plasma parameters.  A brief concluding 
 remark is made in Sec. IV.
\section{Theoretical model}\label{sec:theory}
In the standard fluid model treatment \cite{rao, shukla_book} of a dusty plasma for studying low frequency phenomena
 ($\omega \sim \omega_{pd}\ll\omega_{pi}<\omega_{pe}$, where $\omega_{pd}$, $\omega_{pi}$ and $\omega_{pe}$ are the dust, ion and electron plasma frequencies respectively) in the regime where the dust dynamics is important, it is appropriate to treat
the electrons and the ions as light  fluids that can be described by Boltzmann distributions
and to use the full set of hydrodynamic equations (momentum, continuity and Poisson equations) to describe the dynamics of the dust component. Hence in this paper, the electrons and the ions 
are treated as inertialess and are assumed to be in local thermodynamical 
equilibrium with their number density obeying the Boltzmannian distribution. Thus the densities of electrons and ions at temperature $T_e$ and $T_i$ can be written in a normalized form as \cite{rao, shukla_book}
\begin{eqnarray}
\nonumber
&n_e&=\exp(\sigma\phi),\label{eqn:e_density1}\\
&n_i&=\exp(-\phi). \label{eqn:i_density2}
\end{eqnarray}

\noindent
where  $\sigma$ is the ratio of ion temperature and electron temperature, $n_e$ and $n_i$ have been normalized by their equilibrium values $n_{e0}$ and $n_{i0}$ respectively and $\phi$ has been normalized as $\phi = e\phi / k_B T_{i}$. Here $e$ and $k_B$ denote the electronic charge and the Boltzmann constant, respectively.
To describe the dynamics of dust particles, we use  the well known Generalized Hydrodynamic model \cite{kaw} that takes into account strong coupling effects in a 
phenomenological manner by introducing visco-elastic effects and a modified compressibility. In the regime where existence of solitonic waves and their propagation characteristics are important the predominant  change due to strong coupling effects are in the dispersion properties. This is manifested through a change in the compressibility as seen in the linear effect of a turnover in the dispersion relation. Accordingly, for our solitonic study we retain the compressibility effect and neglect dissipative effects arising from viscosity and dust neutral collisions. These dissipative effects when important would cause a damping of the solitary pulse. The neglect of dissipative effects is a valid approximation in the so called ``kinetic regime'' when $\omega \tau_m >> 1$ where $\omega$ is the mode freqency and $\tau_m$ is the relaxation (memory) time. Thus our model fluid equations for the dust component consisting of the continuity and the momentum coupled to the Poisson equation can be written as,
\begin{eqnarray} 
\frac{\partial n_d}{\partial t}& + & \frac{\partial (n_d v_d)}{\partial x} = 0,\label{eqn:con}\\ \nonumber\\
\frac{\partial v_d}{\partial t}& + & v_d\frac{\partial v_d}{\partial x} = \frac{\partial \phi}{\partial x}-\frac{\mu^\prime}{n_d}\frac{\partial n_d}{\partial 
x} \;\hspace*{0.1 in}\text{and} \label{eqn:mom}\\ \nonumber\\
\frac{\partial^ 2\phi}{\partial x^2} & = & (n_d+\mu_e n_e-\mu_i n_i).\label{eqn:pos}
\end{eqnarray}
This model has already been implemented successfully to explain the turn over of dispersion relation of DAWs \cite{pintu_a} and the existence of shear waves \cite{jyoti_t, pintu_t} in dusty plasma. 
The contribution due to the compressibility ($\mu$) in the momentum equation (Eq.~(\ref{eqn:mom})) is expressed in terms of  $\mu^\prime$, where $\mu^\prime = \frac{\mu T_d}{Z_d T_i}$,  where $T_d$ denotes the dust temperature. Following \cite{kaw} we can define the compressibility as,
\begin{eqnarray}
\mu  = \frac{1}{T_{d}}\left (\frac{\partial P}{\partial n}\right)_{T_{d}} = 1+\frac{u(\Gamma)}{3}+\frac{\Gamma} {9}\frac{\partial 
u(\Gamma)}{\partial \Gamma}\label{eqn:mu}
\end{eqnarray}
where $u(\Gamma)$ is the free energy of the system and can be expressed as \cite{slattery} 
\begin{eqnarray}
\hspace*{-0.5in} u(\Gamma) = -0.89\Gamma + 0.95\Gamma^{1/4} + 0.19\Gamma^{-1/4}-0.81. \label{eqn:gam}
\end{eqnarray}
In the weakly coupled gaseous phase ($\Gamma\ < 1 $),  $\mu$ is positive but can become negative as $\Gamma$ increases and one gets into the liquid
state. The change in sign of $\mu$ is responsible for the turnover effect in the linear dispersion relation of the dust acoustic wave. 
\par
In the above equations (\ref{eqn:con}--\ref{eqn:pos}), $n_d$ and  $v_d$ represent the normalized number density and the velocity of the dust
fluid, respectively. $m_d$ denotes the mass of dust particles. $\mu_e$ and $\mu_i$ are defined as, $\mu_e = \frac{n_{e0}}{Z_{d0}n_{d0}} $ 
and  $\mu_i = \frac{n_{i0}}{Z_{d0}n_{d0}}.$  By using the quasi-neutrality condition, $n_{e0}=n_{io}-Z_{d0}n_{d0}$, we can write $\mu_e= 1/(\delta-1)$ and $\mu_i= \delta/(\delta-1)$, where $\delta$ is the ratio of equilibrium ion density to electron density. The variables $t$, $x$, $n_d$ and $u_d$ are normalized by $w_d^{-1}= (m_d/4 \pi n_{d0}Z_{d0}^2e^2)^{1/2}$, $\lambda_d = (k_BT_i/4\pi Z_{d0}n_{d0}e^2)^{1/2}$, $n_{d0}$ and $C_d = (Z_{d0}k_BT_i/m_d)^{1/2}$ respectively, where $n_{d0}$ is the unperturbed number density of the dust particle and $Z_{d0}$ is the unperturbed number of electrons residing on the dust 
 particles. We do not consider any charge fluctuation  of dust fluid in our model.
 \par Now we consider the excitation of two solitary waves A and B that are far apart from each other and let them 
propagate towards each other such that after sometime they interact and make a head-on collision. We consider the regime where the 
perturbations are small enough for the weakly nonlinear approximation to hold good. 
 We therefore expect the collision to be quasielastic leading to shifts of the post collision trajectories (phase shift).
 Here we are interested to study the dynamics of these solitary waves in presence of strong coupling effect.
 In order to analyse the effect of collision, we employ an extended PLK perturbation method \cite{jef,huang}.
 This PLK  method is a combination of the standard reductive perturbation method \cite{ch,mmm} with the technique of strained 
coordinates. The main idea of this perturbation method is as follows. In the limit of long wavelengh approximation, asymptotic expansions
 for both the flow field variables and spatial or time coordinates are used. This makes a uniformly valid asymptotic expansion
 (i.e., phase shifts) of the solitary waves after the collision. According to this method, we introduce the stretched coordinates
 \color{black}
\begin{eqnarray}
\xi &=& \epsilon(x-\lambda t)+\epsilon^ 2 P_0(\eta , \tau)+\epsilon^ 3 P_1(\xi , \eta , \tau)+..., \label{eqn:jhi}\\ \nonumber\\
\eta &=& \epsilon(x+\lambda t)+\epsilon^ 2  Q_0(\xi , \tau)+\epsilon^ 3Q_1(\xi , \eta , \tau)+..., \label{eqn:ita}\\ \nonumber\\
\tau &=& \epsilon^ 3 t.\label{eqn:tau}
\end{eqnarray} 
Where $\xi$ and $\eta$ denote the space coordinates of the trajectories of the two solitons travelling to the right and left, respectively. We are assuming that 
solitons have small amplitude $\sim \epsilon$ (where $\epsilon$ is a formal smallness (perturbation) parameter characterizing the strength
of nonlinearity). The wave velocity
$\lambda$ and the variables $P_j$ and $Q_j$ are to be determined (where $j=1, 2, 3 ... $). Using Eq.~(\ref{eqn:jhi})--~(\ref{eqn:tau}), we have
 \begin{eqnarray}
 \frac{\partial}{\partial x}&=& \epsilon \left(\frac{\partial}{\partial \xi}+\frac{\partial}{\partial \eta}\right)+
\epsilon^3\left(P_{0\eta}\frac{\partial}{\partial \xi}+Q_{0\xi}\frac{\partial}{\partial \eta}\right)+..., \label{eqn:diff}\\ \nonumber\\
\frac{\partial}{\partial t}&=& \epsilon \lambda \left(-\frac{\partial}{\partial \xi}+
\frac{\partial}{\partial \eta}\right)\nonumber\\&&\hspace*{0.05 in}+\epsilon^3\left(\frac{\partial}{\partial\tau}+
\lambda P_{0\eta}\frac{\partial}{\partial \xi}-\lambda Q_{0\xi}\frac{\partial}{\partial \eta}\right)
+..., \label{eqn:diff1}\\\nonumber\\
\frac{\partial^2}{\partial x^2}&=&\epsilon^2\left(\frac{\partial}{\partial\xi}+\frac{\partial}{\partial \eta}\right)^2
\nonumber\\&&+\epsilon^4\left(\frac{\partial}{\partial\xi}+\frac{\partial}{\partial \eta}\right)
\left(P_{0\eta}\frac{\partial}{\partial \xi}+Q_{0\xi}\frac{\partial}{\partial \eta}\right)
\nonumber\\&&+\epsilon^4\left(P_{0\eta}\frac{\partial}{\partial \xi}+Q_{0\xi}\frac{\partial}{\partial \eta}\right)
\left(\frac{\partial}{\partial\xi}+\frac{\partial}{\partial \eta}\right)+... .\label{eqn:diff2}
\end{eqnarray}
Where $P_{0\eta}=\partial P_0/ \partial \eta$ and $Q_{0\xi}=\partial Q_0/ \partial \xi$. The asymptotic expansions of perturbed quantities ($n_d, v_d$ and $\phi$ ) are given as:
\begin{eqnarray}
&n_d& = 1+ \epsilon^ 2n_1+\epsilon^ 3n_2+...,\label{eqn:diff asymnd}\\\nonumber\\
&u_d &= \epsilon^ 2u_1+\epsilon^ 3u_2+...,\label{eqn:asymvd}\\\nonumber\\
&\phi& =\epsilon^ 2\phi_ 1+\epsilon^ 3\phi_ 2+....\label{eqn:asymphi}
\end{eqnarray}
Substituting Eqs.~(\ref{eqn:diff})--(\ref{eqn:asymphi}) into Eqs.~(\ref{eqn:con})--(\ref{eqn:pos}), and equating the quanitities with
equal power of $\epsilon$, we obtain a set of coupled equations at different orders of $\epsilon$. The leading order term (at order $\epsilon^2$) gives, 
\begin{eqnarray}
\lambda\left(-\frac{\partial}{\partial \xi}+\frac{\partial}{\partial \eta}\right)n_1&+&
\left(\frac{\partial}{\partial \xi}+\frac{\partial}{\partial \eta}\right)u_1 =0,\label{eqn:leadconti}\\ \nonumber\\
\lambda\left(-\frac{\partial}{\partial \xi}+\frac{\partial}{\partial \eta}\right)u_1
&=& \left(\frac{\partial}{\partial \xi}+\frac{\partial}{\partial \eta}\right)\phi_1 \nonumber\\
&-&\mu^\prime\left(\frac{\partial}{\partial \xi}+\frac{\partial}{\partial \eta}\right)n_1,\label{eqn:leadv}\\\nonumber\\
n_1 &=& -Q\phi_1\label{eqn:leadph}
\end{eqnarray}
where, $Q\left(= \mu_e\sigma+\mu_i\right)$ is an important parameter related to the density and temperature ratio of ion and electrons.
Solving Eqs.~(\ref{eqn:leadconti})--(\ref{eqn:leadph}) we get,
\begin{eqnarray}
 \phi_1 &=& \Phi_1\left(\xi,\tau\right)+\Phi_2\left(\eta,\tau\right),\label{eqn:solph}\\ \nonumber \\
 n_1 &=& -Q\left(\Phi_1\left(\xi,\tau\right)+\Phi_2\left(\eta,\tau\right)\right),\label{eqn:solden}\\ \nonumber \\
 u_1 &=& -\frac{(1+Q\mu^\prime)}{\lambda}\left(\Phi_1 (\xi , \tau)-\Phi_2(\eta , \tau)\right)\label{eqn:solvelo}
\end{eqnarray}
and with the solvability condition \textit{i.e.}, the condition to obtain a uniquely defined $n_1$ and $u_1$ from Eqs.~(\ref{eqn:leadconti})--(\ref{eqn:leadph}) when $\phi_1$ is given by Eq.~(\ref{eqn:solph}), the phase velocity $\lambda=
\sqrt{\frac{\left(1+Q\mu^\prime\right)}{Q}}$ is also obtained. The unknown functions $\Phi_1$ and $\Phi_2$ will be determined at higher 
order.
Eqs.~(\ref{eqn:solph})--(\ref{eqn:solvelo}) imply that, at the leading order, we have two waves, one of which, $\Phi_1\left(\xi,\tau\right)$,
 is travelling right, and the other one, $\Phi_2\left(\eta,\tau\right)$, is travelling left. At the next order (i.e., $\epsilon^3$),
 we have a system of equation given as:
 \begin{eqnarray}
\lambda\left(-\frac{\partial}{\partial \xi}+\frac{\partial}{\partial \eta}\right)n_2&+&
\left(\frac{\partial}{\partial \xi}+\frac{\partial}{\partial \eta}\right)u_2 =0\label{eqn:leadcon},\\ \nonumber\\
\lambda\left(-\frac{\partial}{\partial \xi}+\frac{\partial}{\partial \eta}\right)u_2
&=& \left(\frac{\partial}{\partial \xi}+\frac{\partial}{\partial \eta}\right)\phi_2 \nonumber\\
&-&\mu^\prime\left(\frac{\partial}{\partial \xi}+\frac{\partial}{\partial \eta}\right)n_2,\label{eqn:leadvd}\\\nonumber\\
n_2 &=& -Q\phi_2.\label{eqn:leadphi}
\end{eqnarray}
These are similar to the leading order equations, so that, the solutions also have the following shape:
\begin{eqnarray}
 \phi_2 &=& \Psi_1\left(\xi,\tau\right)+\Psi_2\left(\eta,\tau\right),\label{eqn:solphi}\\ \nonumber \\
 n_1 &=& -Q\left(\Psi_1\left(\xi,\tau\right)+\Psi_2\left(\eta,\tau\right)\right),\label{eqn:soln1}\\ \nonumber \\
 u_1 &=& -\frac{(1+Q\mu^\prime)}{\lambda}\left(\Psi_1 (\xi , \tau)-\Psi_2(\eta , \tau)\right)\label{eqn:solu1}
\end{eqnarray}
 where $\Psi_1$ and $\Psi_2$ are to be determined. 
 Now from the next leading order (i.e., $\epsilon^4$), we have a system of equations which can be written as,
 \begin{eqnarray}
\lambda\left(-\frac{\partial}{\partial \xi}+\frac{\partial}{\partial \eta}\right)n_3+\left(\frac{\partial}{\partial\tau}+
\lambda P_{0\eta}\frac{\partial}{\partial \xi}-\lambda Q_{0\xi}\frac{\partial}{\partial \eta}\right)n_1\nonumber\\
+\left(\frac{\partial}{\partial \xi}+\frac{\partial}{\partial \eta}\right)u_3
+\left(\frac{\partial}{\partial \xi}+\frac{\partial}{\partial \eta}\right)n_1u_1\nonumber\\
+ \left(P_0\frac{\partial}{\partial\xi}+Q_0\frac{\partial}{\partial\eta}\right)u_1=0,\nonumber\\
\end{eqnarray}
%
\begin{eqnarray}
\lambda\left(-\frac{\partial}{\partial \xi}+\frac{\partial}{\partial \eta}\right)u_3+\left(\frac{\partial}{\partial\tau}+
\lambda P_{0\eta}\frac{\partial}{\partial \xi}-\lambda Q_{0\xi}\frac{\partial}{\partial \eta}\right)u_1\nonumber\\
+u_1\left(\frac{\partial}{\partial \xi}+\frac{\partial}{\partial \eta}\right)u_1=
\left(\frac{\partial}{\partial \xi}+\frac{\partial}{\partial \eta}\right)\phi_3\nonumber\\
+ \left(P_0\frac{\partial}{\partial\xi}+Q_0\frac{\partial}{\partial\eta}\right)\phi_1
-\mu^\prime \left(\frac{\partial}{\partial \xi}+\frac{\partial}{\partial \eta}\right)n_3\nonumber\\
+\mu^\prime n_1\left(\frac{\partial}{\partial \xi}+\frac{\partial}{\partial \eta}\right)n_1
-\mu^\prime\left(P_0\frac{\partial}{\partial\xi}+Q_0\frac{\partial}{\partial\eta}\right)n_1,\nonumber\\
\end{eqnarray}
\begin{equation}
 \left(\frac{\partial}{\partial \xi}+\frac{\partial}{\partial \eta}\right)^2\phi_1
 = n_3+Q\phi_3+R/2\phi_1^2\label{eqn:n3}
\end{equation}
where $R= \mu_e\sigma^2-\mu_i$.
Solving above Eqs.~(\ref{eqn:leadconti})--(\ref{eqn:n3}) we can find 
\begin{eqnarray}
 \lambda\frac{\partial^2u_3}{\partial\xi\partial\eta} = \frac{d}{2\lambda}\frac{\partial}{\partial\xi}
\left[\frac{\partial\Phi_1}{\partial\tau}+ a \Phi_1 \frac{\partial \Phi_1}{\partial \xi}+
b\frac{\partial^3\Phi_1}{\partial\xi^3} \right]\nonumber\\
+\frac{d}{2\lambda}\frac{\partial}{\partial\eta}
\left[\frac{\partial\Phi_2}{\partial\tau}- a \Phi_2 \frac{\partial \Phi_2}{\partial \eta}-
b\frac{\partial^3\Phi_2}{\partial\eta^3} \right]\nonumber\\ +\left[dP_{0\eta}+c \Phi_2\right]\frac{\partial^2\Phi_1}{\partial\xi^2}-
\left[dQ_{0\xi}+c \Phi_1\right]\frac{\partial^2\Phi_2}{\partial\eta^2}.\label{eqn:kdv}
\end{eqnarray}
Integrating the above equation with respect to the variables $\xi$ and $\eta$ yields
\begin{eqnarray}
\lambda u_3 &=& \frac{d}{2\lambda}\int\left(\frac{\partial\Phi_1}{\partial\tau}+ a \Phi_1 \frac{\partial \Phi_1}{\partial \xi}+
b\frac{\partial^3\Phi_1}{\partial\xi^3} \right)d\eta\nonumber\\&+&
\frac{d}{2\lambda}\int\left(\frac{\partial\Phi_2}{\partial\tau}- a \Phi_2 \frac{\partial \Phi_2}{\partial \eta}-
b\frac{\partial^3\Phi_2}{\partial\eta^3} \right)d\xi\nonumber\\&&\hspace*{-0.5 in}+
\int\int\left(d\frac{\partial P_0}{\partial\eta}+c \Phi_2\right)d\xi d\eta
-\int\int\left(d\frac{\partial Q_0}{\partial\xi}+c \Phi_1\right)d\xi d\eta,\nonumber\\\label{eqn:kdv1}
\end{eqnarray}
where the constants are defined as, $a=\frac{1}{2d}\left(-\frac{3d^2}{\lambda}+\mu^\prime Q^2\lambda-\frac{R\lambda}{Q}\right), b=
\frac{\lambda}{2Qd},
c=\frac{-1}{4\lambda}\left(\frac{R\lambda}{Q}-\mu^\prime Q^2\lambda-\frac{d^2}{\lambda}\right)$ and $d=\left(1+Q\mu^\prime\right)$.\par
 The first (second) term in the Eq.~(\ref{eqn:kdv1}) will be proportional to $\eta(\xi)$ because the integrated function is independent
of $\eta(\xi)$. Thus the first two terms of Eq.~(\ref{eqn:kdv1}) are 
secular terms, which must be eliminated in order to avoid
spurious resonances. Hence we have
\begin{eqnarray}
 \frac{\partial\Phi_1}{\partial\tau}+ a \Phi_1 \frac{\partial \Phi_1}{\partial \xi}+
b\frac{\partial^3\Phi_1}{\partial\xi^3}=0,\label{eqn:k}\\
\frac{\partial\Phi_2}{\partial\tau}- a \Phi_2 \frac{\partial \Phi_2}{\partial \eta}-
b\frac{\partial^3\Phi_2}{\partial\eta^3}=0.\label{eqn:m}
\end{eqnarray}
The third and fourth terms in Eq.~(\ref{eqn:kdv1}) are not secular terms in this order, but they will become secular in the next
order \cite{jef}. Hence we have
\begin{eqnarray}
 \frac{\partial P_0}{\partial\eta}=-\frac{c}{d} \Phi_2,\label{eqn:secu}\\
 \frac{\partial Q_0}{\partial\xi}=-\frac{c}{d}  \Phi_1.\label{eqn:secu1}
\end{eqnarray}
Eqs.~(\ref{eqn:k}) and~(\ref{eqn:m}) are two-side travelling wave KdV equations in the reference frames of $\xi$
and $\eta$, respectively. Their corresponding solutions are given by,
\begin{equation}
 \Phi_1 = \Phi_A  \text{sech}^ 2 \left[\left(\frac{a\Phi_A}{12b}\right)^{1/2}\left(\xi-\frac{1}{3}a\Phi_A\tau\right)\right],\label{eqn:phia}
\end{equation}
\begin{equation}
 \Phi_2 = \Phi_B  \text{sech}^ 2\left[\left(\frac{a\Phi_B}{12b}\right)^{1/2}\left(\eta+\frac{1}{3}a\Phi_B\tau\right)\right]\label{eqn:phib}
\end{equation}
where $\Phi_A=3\delta M_1/a$ and $\Phi_B=3\delta M_2/a$ are the amplitudes of the two solitons A and B in their initial positions. The 
leading phase changes
due to collision can be calculated from Eqs.~(\ref{eqn:secu})--(\ref{eqn:phib}) and given as:
\begin{eqnarray}
 P_0 &=& -\frac{c}{d}\left(\frac{12b\Phi_B}{a}\right)^{1/2}\nonumber\\
 &\times&\left\{\text{tanh} \left[\left(\frac{a\Phi_B}{12b}\right)^{1/2}\left(\eta+
 \frac{1}{3}a\Phi_B\tau\right)\right]+1\right\},
\end{eqnarray}
\begin{eqnarray}
 Q_0 &=& -\frac{c}{d}\left(\frac{12b\Phi_A}{a}\right)^{1/2}\nonumber\\
 &\times&\left \{\text{tanh} \left[\left(\frac{a\Phi_A}{12b}\right)^{1/2}\left(\xi-
 \frac{1}{3}a\Phi_A\tau\right)\right]-1\right\}.
\end{eqnarray}
Hence, up to $O\left(\epsilon^2\right)$, the trajectories of the two solitary waves for weak head-on interactions in
presence of strong coupling effect can be written as:
\begin{eqnarray}
\xi &=& \epsilon\left(x-\lambda t\right)-\epsilon^2 \frac{c}{d}\left(\frac{12b\Phi_B}{a}\right)^{1/2}\times\nonumber\\
&&\hspace*{-0.5 in}\left\{\text{tanh} \left[\left(\frac{a\Phi_B}{12b}\right)^{1/2}
 \left(\eta+\frac{1}{3}a\Phi_B\tau\right)\right]+1\right\}+O\left(\epsilon^3\right), \label{eqn:jhiv}
\end{eqnarray}
\begin{eqnarray}
\eta &=& \epsilon\left(x+\lambda t\right)-\epsilon^2 \frac{c}{d}\left(\frac{12b\Phi_A}{a}\right)^{1/2}\times\nonumber\\
&&\hspace*{-0.5 in}\left\{\text{tanh} \left[\left(\frac{a\Phi_A}{12b}\right)^{1/2}
 \left(\xi-\frac{1}{3}a\Phi_A\tau\right)\right]-1\right\}+O\left(\epsilon^3\right).\label{eqn:itav}
\end{eqnarray}
\par To obtain the phase shifts due to a head-on collision of the two solitons, we assume that the solitons A and B are,
asymptotically far from each other at the initial time $\left(t=-\infty\right)$, i.e., soliton A is at $\left(\xi=0, \eta=-\infty\right)$
and soliton B is at $\left(\eta=0, \xi=+\infty\right)$. After  the collision $\left(t=+\infty\right)$, solitons A is far to the right of
soliton B, i.e., soliton A is at $\left(\xi=0, \eta=+\infty\right)$ and soliton B is at $\left(\eta=0, \xi=-\infty\right).$ Using Eqs.~(\ref{eqn:jhiv}) and~(\ref{eqn:itav}) we obtain the corresponding phase shifts $\Delta A$ and $\Delta B$ as follows:
\begin{eqnarray}
 &\Delta A&=\epsilon\left(x-\lambda t\right)|_{\xi=0,\eta=+\infty}-\epsilon\left(x-\lambda t\right)|_{\xi=0,\eta=-\infty},\nonumber\\
 &\Delta B&=\epsilon\left(x+\lambda t\right)|_{\eta=0,\xi=-\infty}-\epsilon\left(x+\lambda t\right)|_{\eta=0\xi=+\infty}.\nonumber
\end{eqnarray}
This gives the phase shift in solitons A and B which can be expressed as:
\begin{eqnarray}
 \Delta A=2\epsilon^2\frac{c}{d}\left(\frac{12b\Phi_B}{a}\right)^{1/2},\label{eqn:phaseA}\\
 \Delta B=-2\epsilon^2\frac{c}{d}\left(\frac{12b\Phi_A}{a}\right)^{1/2}.\label{eqn:phaseB}
\end{eqnarray}
\begin{figure}[ht]
\includegraphics[width=0.5\textwidth]{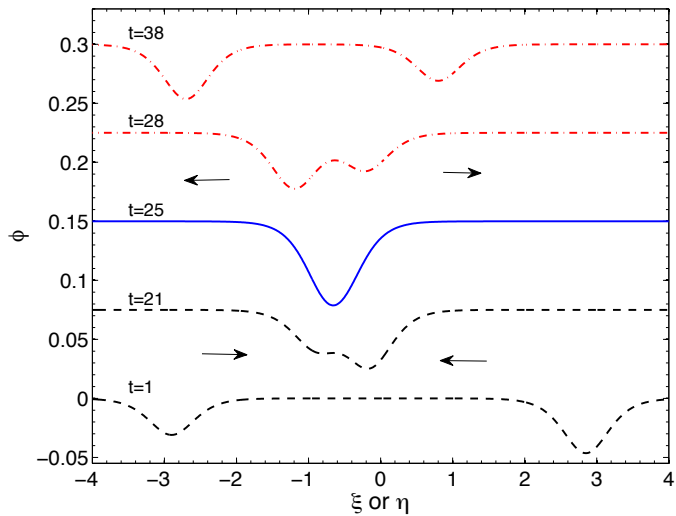}
\caption{ (Color online) Time evolution of two counter-propagating solitons for  $\sigma=0.2$, $\epsilon = 0.1$,
$\delta =1.031$ and $\mu=-20$. Different colour corresponds to three different stages (before collision, during collision and after collision) of propagation. } 
\label{fig:fig1} 
\end{figure}

It should be mentioned here that since the PLK perturbation technique makes use of the decomposition,  
$\phi_1=\Phi_1\left(\xi,\tau\right)+\Phi_2\left(\eta,\tau\right)$ (see Eq.~\ref{eqn:solph}), of wave potential fluctuations amounting to a linear 
superposition of KdV solitons (Eq.~\ref{eqn:phia} and Eq.~\ref{eqn:phib}), it may not provide an accurate description of the amplitude dynamics during the collision process when the two solitons overlap. However the resultant phase shift is quite accurately determined within the perturbation limits since it involves an aymptotic calculation with the solitons well separated from each other. This fact has been well demonstrated and the technique successfully employed by several past authors studying head on collision of two solitary waves in various media \cite{xue, ghosh, versh, labany}. For a detailed critique of the strengths and limitations of the method the reader is referred to the paper by Verheest et al. \cite{versh}. \\
Our main objective in this work is to assess the influence of strong coupling on these phase shifts and in the next section we present our results obtained using the above analytic relations (Eq.~\ref{eqn:phaseA} and Eq.~\ref{eqn:phaseB}) for various plasma parameters.\\ 
\section{Results and Discussions}\label{sec:results}
$\Phi_1 (\xi, \tau)$ and $\Phi_2(\eta, \tau)$  are evaluated by solving the above KdV equations  (Eq.~(\ref{eqn:k}) and 
(\ref{eqn:m})). The time evolution of these two solitary waves which propagate towards each other can be plotted by 
replacing  $\Phi_1$ and $\Phi_2$ in Eq.~(\ref{eqn:solph}).   Fig.~\ref{fig:fig1} shows the time evolution of two counter 
propagating solitons travelling with different amplitudes and widths. It is noticed that the DA solitary wave with higher amplitude  travels faster than that of smaller 
amplitude which was also observed experimentally \cite{Pintu}. It is worth mentioning that we have used $\epsilon = 0.1$ in our 
calculations similar to Xue \textit{et al.}\cite{xue}.\par
The waves which are coming towards each other,  penetrate and slightly dip immediately after the collision and return to their 
initial amplitudes at a later time. 
\begin{figure}[hb]
\includegraphics[width=0.48\textwidth]{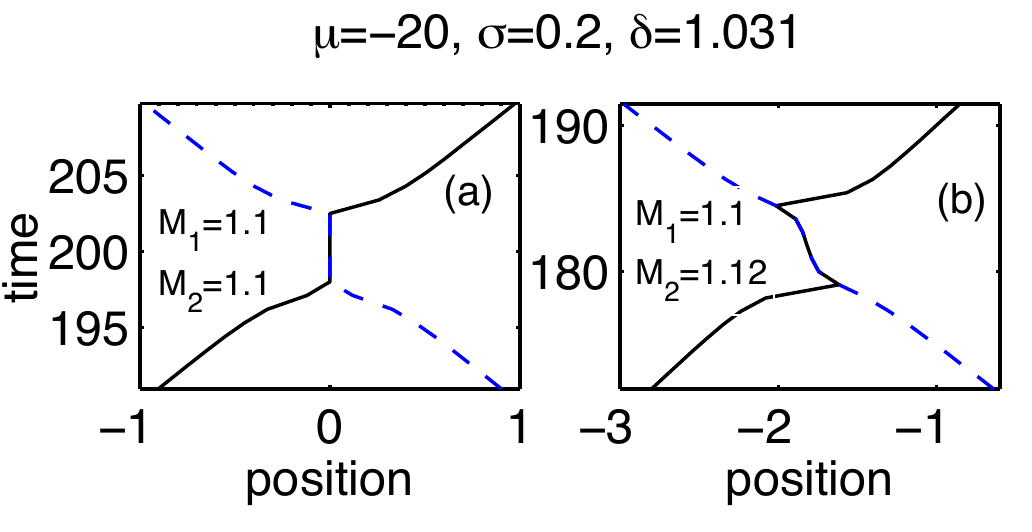}
\caption{(Color online) Trajectory of solitons of (a) same ($M_1=M_2=1.1$) and (b) different ($M_1=1.1$ and $M_2=1.12$) amplitudes/Mach numbers. The solid and dashed lines represent the trajectory of soliton A and soliton B, respectively.  The dusty plasma parameter are taken as  $\mu=-20$, $\sigma=0.2$  and $\delta=1.031$.}
\label{fig:fig5} 
\end{figure}
Fig.~\ref{fig:fig5} shows the trajectories of two counter propagating solitons with identical (see Fig.~\ref{fig:fig5}(a)) and with
different Mach
numbers (see Fig.~\ref{fig:fig5}(b)) respectively. It is to be noted that different Mach numbers imply different amplitudes with 
higher Mach numbers corresponding to larger amplitudes. As can be seen the phase diagram 
changes significantly when the amplitudes differ 
from each other. It is clear from Fig.~\ref{fig:fig5}(b) that the soliton with larger amplitude (soliton B) forces the 
other soliton with smaller amplitude (soliton A) to take a longer time to recover its shape after the collision. Hence it can be 
concluded that the phase shift of the smaller soliton is comparatively larger than that of the bigger one. \par 
Since soliton A is travelling to the right and soliton B is travelling to the left, it is seen from Eqs.~(\ref{eqn:phaseA}) 
and~(\ref{eqn:phaseB}) that each soliton has a positive (or negative) phase shift in its traveling direction due to the collision. 
The sign of phase shift depends on the dusty plasma parameters mainly on $\delta$ and $\mu$ which will be  discussed in more details 
later in this section. \par 
It is clearly seen from Eqs.~(\ref{eqn:phaseA}) and (\ref{eqn:phaseB}) that the phase shifts depend on
the dusty plasma parameters (i.e. $\delta,  \mu$ and  $\sigma $) and the initial amplitudes of the two solitary waves 
($\Phi_A ~\text{and} ~\Phi_B)$.  The co-efficient \lq $a$' remains negative whereas the coefficients \lq $b$' and \lq $d$' remain 
positive 
for all the values of above mentioned parameters. But the co-efficient \lq $c$' changes its sign for a particular set of these
parameters.  Since, the phase shift is directly proportional to \lq{c}',  its sign changes with the sign of \lq c'. A negative phase 
shift implies that the velocity of each soliton reduces at the time of the head-on collision \cite{li}.  It further signifies that 
they either travel the same distance in a longer time or a shorter distance in the same interval of time.  
\par
\begin{figure}[ht]
\includegraphics[width=0.45\textwidth]{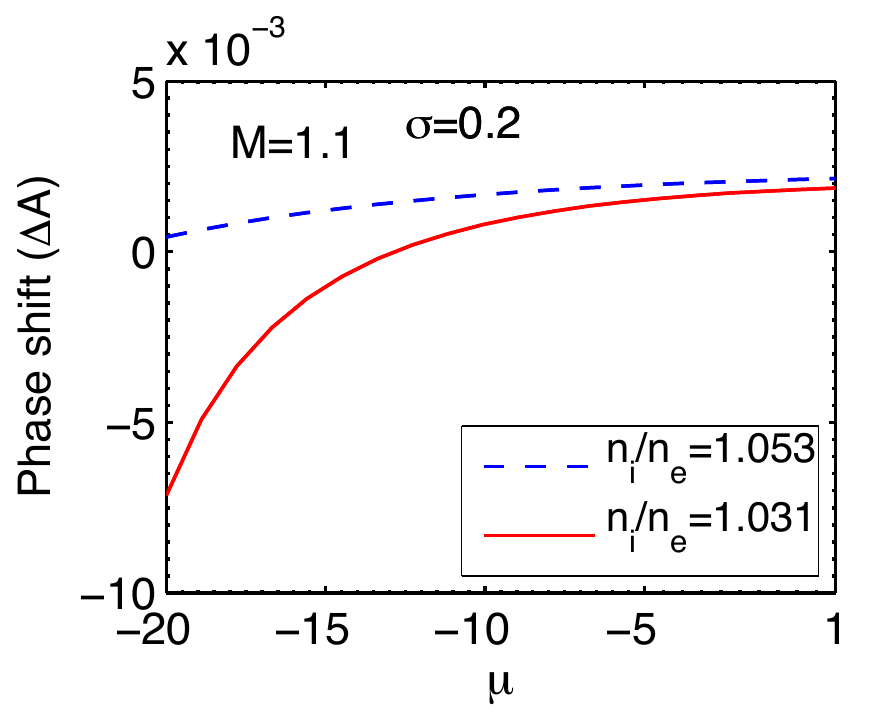}
\caption{(Color online) Variation of phase shift of first soliton ($\Delta A$) with $\mu$ for different values of 
$\delta$ (dashed line and solid line represent $\delta=1.053$ and  $\delta=1.031$, respectively.) at $\sigma=0.2$, $\epsilon = 0.1$, $M=1.1$.}
\label{fig:fig2} 
\end{figure}
The variation of phase shift with compressibility ($\mu$) is shown in Fig.~\ref{fig:fig2} for  $\sigma=0.2$\cite{xue} and $M=1.1$.
 \begin{figure}[hb]
\includegraphics[width=0.4\textwidth]{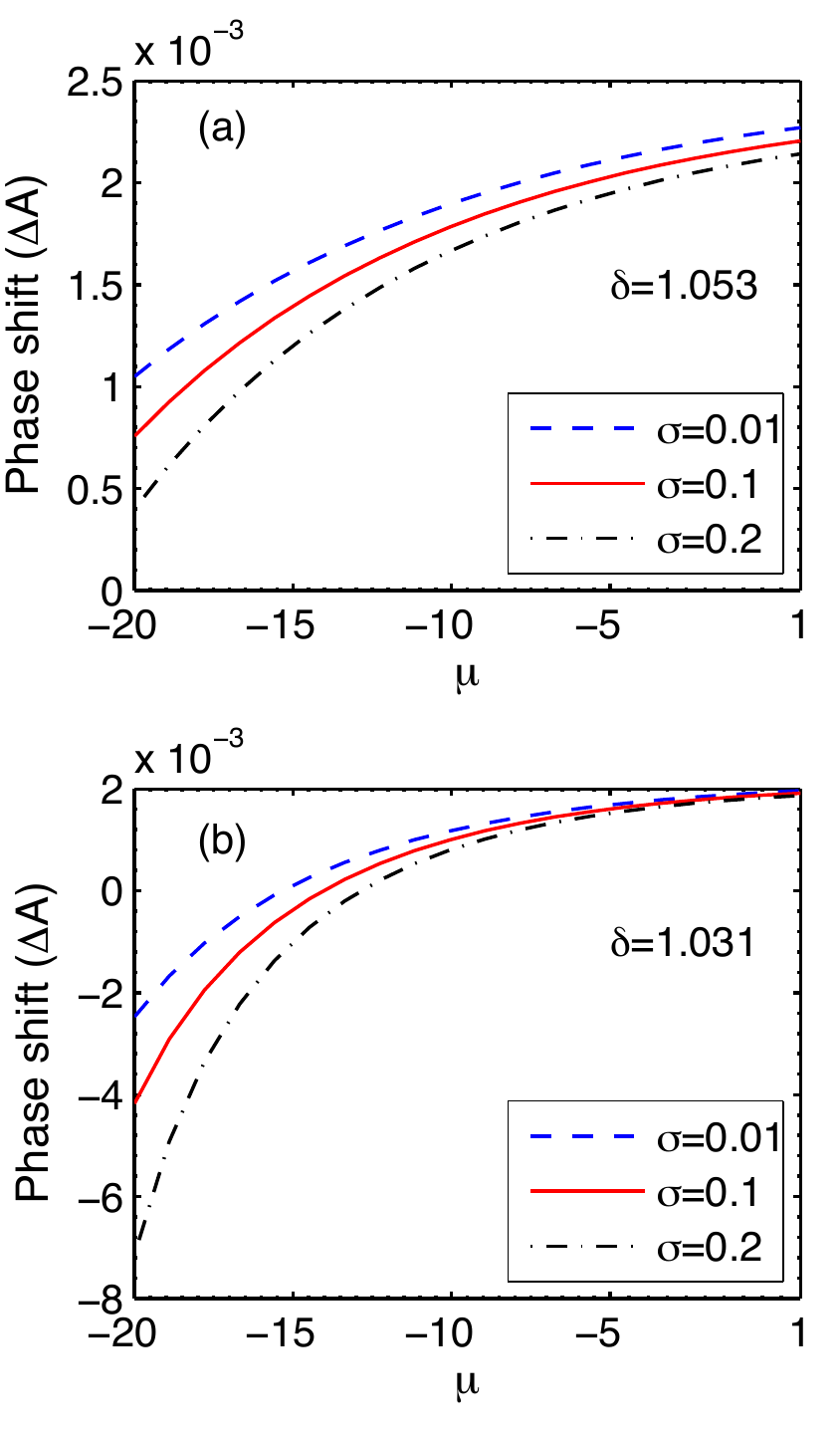}
\caption{(Color online) Variation of phase shift $\Delta A$ with $\mu$ for (a) $\delta=1.053$ and (b) $\delta=1.031$. Dashed, solid and dash-doted lines represent $\sigma=0.01, \sigma=0.1$, and $\sigma=0.2$, respectively.} 
\label{fig:3}
\end{figure}
Our estimates for the phase shifts have been carried out for plasma parameters that are closely related to those of experiments done in the past for solitary waves. For example, in the experiment
on soliton propagation carried out in \cite{Pintu} the typical values of  $\sigma$ and $\delta$ are  $0.0375$ and $1.75$ respectively and they vary around these values with experimental changes of the discharge parameters. In 
another experimental paper by Sharma \textit{et al.} \cite{bailung2014}, the typical values of these parameters are $\sigma=0.02$ and $\delta=1.1$. As stated in their paper, for different pressures, the value of the dust charge number changes and hence the value of 
$\delta$ changes. Our choice of parameter values are in the same range and therefore quite relevant for experimental investigations. 
Additionally, we have also varied the coupling parameter from the weakly coupled regime ($\Gamma<1$) to strongly coupled regime 
($\Gamma>1$) to get the value of $\mu$ from the expression of the compressibility used in the manuscript (Eq.~\ref{eqn:mu} and Eq.~\ref{eqn:gam}). Our choice of the range of the variation of 
the coupling parameter is also close to that reported in many experimental papers \cite{goree1, goree2}. \par
\begin{figure}[ht]
\includegraphics[width=0.4\textwidth]{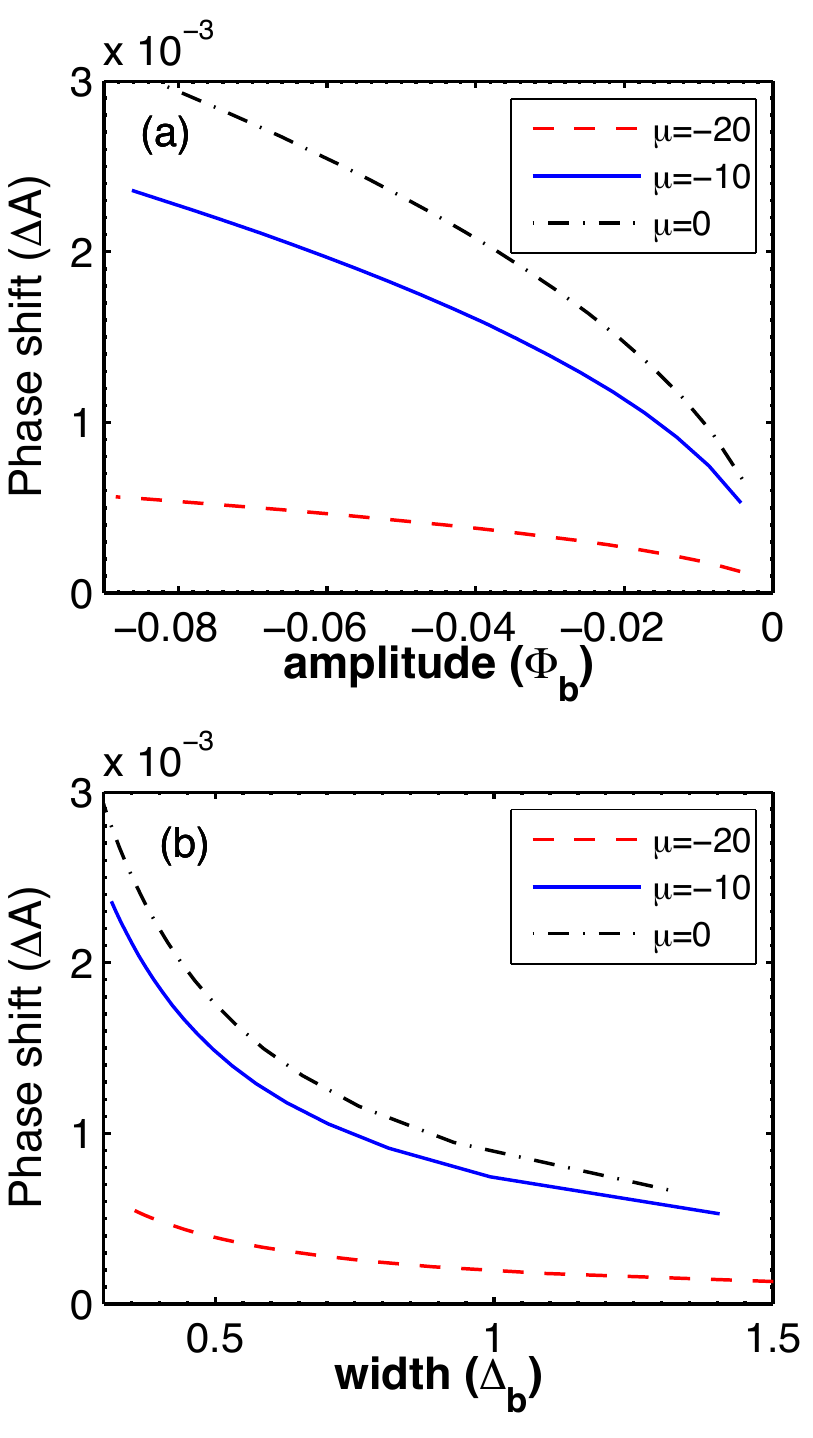}
\caption{(Color online) Variation of phase shift $\Delta A$ with (a) amplitude($\Phi_A$) and (b) width($\Delta_s$)
for three different values  of $\mu$ ($\mu=-20$ (dashed), $\mu=-10$ (solid) and $\mu=0$ (dash-dotted)) at $\sigma=0.2$ and $\delta=1.053$.}
\label{fig:fig4} 
\end{figure}
To study the variations of phase shifts, we have plotted the phase shift of first solitary wave ($\Delta A$) against this parameters. It is 
not necessary to study the phase shift of the second soliton separately as it always shows the  same trend as of the first one but
with negative sign because it is travelling in the opposite direction. The solid and dashed lines depict for $\delta=n_i/n_e=1.031$ 
and  $\delta=n_i/n_e=1.053$, respectively. It is clear from this figure that the phase shift changes significantly in strongly 
coupled regime ($\mu \le 1$) compared to the weakly coupled regime ($\mu \ge 1$) for both the cases. Additionally,  it is also seen
that the phase shift ($\Delta A$) changes its sign for $\delta=1.031$ nearly at $\mu=-13$ for given dusty plasma parameters. It 
suggests that for both the cases, the velocity of soliton A  reduces during collision because of higher 
rigidity of the medium that increases with the decrease of compressibility. \par
We have plotted the variation of phase shift with $\mu$ for $\sigma = $ 0.01 (dashed line), 0.1 
(solid line) and 0.2 (dash-dotted line) in Fig.~\ref{fig:3}. We have chosen the value of $\delta=1.053$ in Fig.~\ref{fig:3}(a) whereas $\delta=1.031$ is chosen for Fig.~\ref{fig:3}(b). In case of Fig.~\ref{fig:3}(a), the phase shift is monotonically decreasing for each $\sigma$ with the increase of coupling parameter (decreasing $\mu$). But in Fig.~\ref{fig:3}(b), the magnitude of phase shift is initially decreasing (upto $\mu=-13$, as discussed in Fig.~\ref{fig:fig2}) and then it increases again with $\mu$. But for both the figures the velocity of soliton A is decreasing with the increase of the rigidity of the medium during collision. It is also found in both the cases that the phase delay decreases with the increase of temperature ratio, $\sigma$ for a given value of $\delta$.\par 
Fig.~\ref{fig:fig4} shows that the phase shift of soliton A changes significantly with the change of solitary amplitude of 
B and its width for three different values of $\mu$. It means for a given value of $\mu$, larger amplitude (or smaller width) of 
soliton B causes larger delay in the propagation of soliton A. This theoretical findings also supports the experimental results of Harvey \textit{et al.} \cite{Harvey}. For a given value of width (or amplitude) the phase-shift decreases with decrease of $\mu$. Decreasing $\mu$ corresponds to the increase of the rigidity of the medium.
\section{Conclusion}
\label{sec:conclusion}
We have theoretically calculated the leading-order phase shift resulting from a head-on collision between two counter propagating dust acoustic solitary waves in an unmagnetized strongly coupled dusty plasma system. The primary objective was to assess the influence of strong coupling on this nonlinear process. We have used the Generalized Hydrodynamic Equations to model the dust dynamics and accounted for strong coupling induced 
dispersive effects through modifications in the compressibility arising from contributions due to $\Gamma$. The variation of the phase shift as a function of the compressibility is studied. In addition the variation due to parameters like the density ratio ($\delta$), the temperature ratio ($\sigma$) of the plasma species and the initial amplitudes of the solitary waves on the phase delay are also investigated. We find that the phase shift from a head-on collision changes significantly in the strongly coupled regime as compared to the weakly coupled regime. We have also found that as we increase the rigidity of the medium the phase shift changes its sign for a given set of dusty plasma parameters. A negative phase shift suggests that the velocity of the solitary waves decreases at the time of the collision. It is also seen that the phase shift decreases with the increase of temperature ratio of ions to electrons. Further a larger amplitude (or smaller width) soliton causes a larger delay. Our model results 
may serve as interesting signatures of nonlinear manifestations of strong coupling effects that could be looked for in collision experiments in a laboratory set-up. They can also form the basis for further theoretical investigations where the effect of higher order corrections and dissipative contributions can be explored.

\end{document}